# Ice Surface Entropy Induction by Humidity or How Humidity Prompts Freezing

J. L. Pérez-Díaz,[a] M. A. Álvarez-Valenzuela,[b] J. Sánchez-García-Casarrubios[b] and S. Jiménez-López[b]

[a] Present address: Departamento de Teoría de la Señal, Universidad de Alcalá, N-II km 33,600. Alcalá de Henares 28805, Spain.
[b] MAG SOAR S.L., Avda. Europa, 82. Valdemoro 28341, Spain.

**Abstract.** In this work we measured Surface Energy and Freezing Temperature of supercooled water droplets in air. We find that freezing of water droplets is triggered at the water-air interface and that freezing progresses faster on the surface than in the bulk. The Freezing Point of water droplets is strongly depressed by dryness in air and how humidity triggers freezing. Additionally it is shown to be a Surface phenomenon related to a transfer of Entropy from water vapour to the surface of ice.

**Keywords:**

**Supercooling; freezing; water; ice; water-air interface; water surface; ice-air interface.**

**Introduction.**

Water has always been of obvious interest to scientists not only because of its omnipresence and importance in human activity, chemistry, biology and climate but also because of its anomalies such as the high-density and low-density 'structures' of supercooled liquid water[1,2].

Although bulk water has a number of counterintuitive, exciting and decisive properties, its surface is a prodigious source of surprise and it is even more crucial for chemical, meteorological and biological processes[3,4]. This should not really be surprising as surfaces and interfaces of solids are known to present distinct characteristic properties different from those of bulk materials[5].

It seems to be only one molecular layer that has a 'structure' different from that of bulk water[6,7], but it is enough to enhance the rate of certain chemical reactions[8,9] or to nucleate freezing of supercooled water[10].

Freezing of water is of interest to scientists as its many technical applications strongly affect human life –for instance ice accretion on planes, wind turbines or electrical lines-. Although bulk water can be supercooled down to around - 41ºC[11,12] before it homogeneously freezes, inhomogeneous freezing of water is triggered on the surface at much higher temperatures[13].

This clearly superficial characteristic makes us to follow an "Interfacial" approach to this topic. Strictly speaking the surface of water in air must be considered as a water-air interface, whose properties depend on those of both (liquid) water and air.

Maybe the best known physical magnitude characterizing the surface or interface is "Surface Energy", also called "Surface Tension" in the case of a liquid. It is defined as the energy per unit area associated with the existence of the surface of a material. It is well-known that surface tension can be measured macroscopically and mesoscopically. In particular, the pendant drop method, which is used in this work, only requires a small drop of water to measure the surface tension of a liquid[14,15].

To further support the fact that "Interfacial" properties depend on those of both water and air, we demonstrated in previous work[16] that surface tension on deionised water droplets decreases with increasing humidity for several temperatures above 5ºC. It was the first time that the surface property of a liquid was shown to depend on the composition of the gas phase.

For that purpose we designed and built an isothermal chamber to measure the surface tension of a liquid droplet at a fixed temperature and a fixed humidity for air at atmospheric pressure. In this chamber the measured temperature of the air differs by less than 0.04 ºC from that of a water droplet.

In the present work we used the same isothermal chamber and the pending drop method[14-16] to measure the surface tension of super-cooled water droplets and their freezing temperature at several humidities.

**Theoretical model and experimental results.**

Figure 1 below shows the surface tension (or surface energy) measured for supercooled deionised water droplets at -9.0 ºC, -3.0 ºC and +5.0 ºC. Surface tension decreases with humidity in all cases, but the dependence is stronger for +5.0 ºC than for the supercooled droplets.

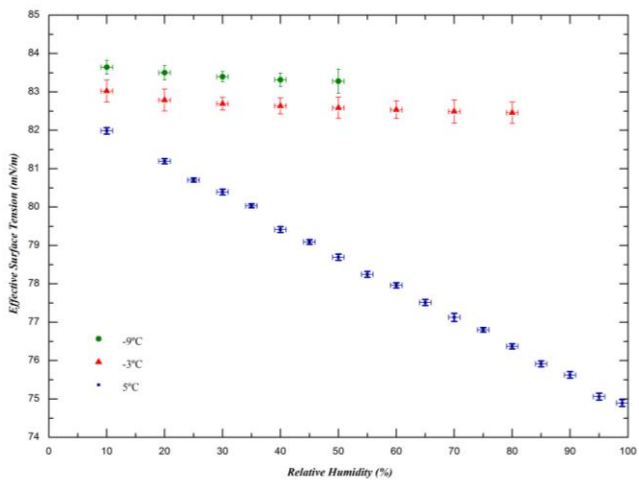

When slowly cooling down the supercooled water droplets, as well as the chamber and the air, in still air in constant relative humidity conditions we found that the lower the relative humidity is, the lower the temperature at which the droplet freezes, as shown in Figure 2. Thus, we can say that dryness 'prevents' freezing of water droplets.

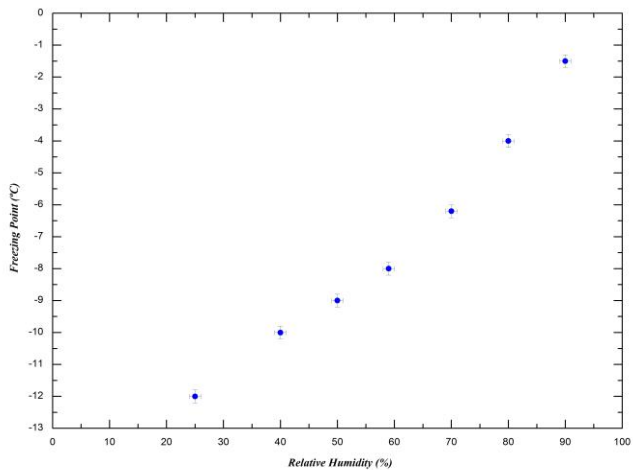

Additionally, isothermal freezing of supercooled water droplets is also triggered when humidity in air is increased just passing over the curve defined in Fig. 2 above.

This dependence of freezing point with humidity in air demonstrates that freezing must be triggered on the surface just interacting with air. In other words, a change of the gas phase induces a phase change in the liquid into solid.

Second evidence can be found in the formation of icicles as shown in fig. 3 below. We used a camera to sequence the freezing process in different conditions of humidity. If freezing started in the volume of the droplet, the outter surface would remain liquid up to the end. In this case the shape of the frozen droplet would remain almost constant –except for merely a homogeneous expansion-. But it is evident from fig. 3 that it is not the case.

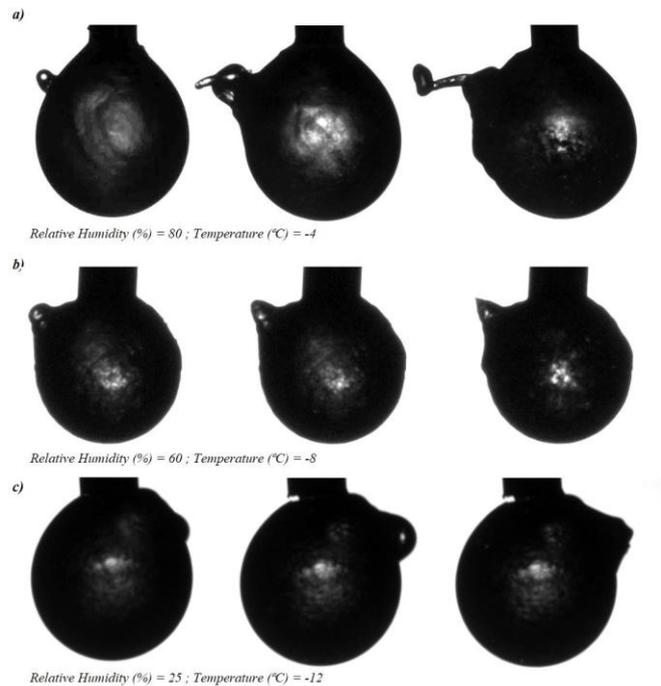

*Relative Humidity (%) = 80 ; Temperature (°C) = -4*

*Relative Humidity (%) = 60 ; Temperature (°C) = -8*

*Relative Humidity (%) = 25 ; Temperature (°C) = -12*

The formation of icicles can be explained as follows. First, freezing of the surface occurs, preventing light from the back reaching the camera. The freezing frontwave progression all along the surface of the droplet is faster than the time between frames (15 ms). Once the surface is frozen, it makes up a sort of rigid shell which still contains liquid supercooled water. Then the progression of the freezing frontwave inwards compresses the liquid water which eventually breaks out through the weakest point of the shell. This generates protruding icicles, as can be seen in Figure 3 above and the following video files (click on each of the following images to follow the Youtube links):

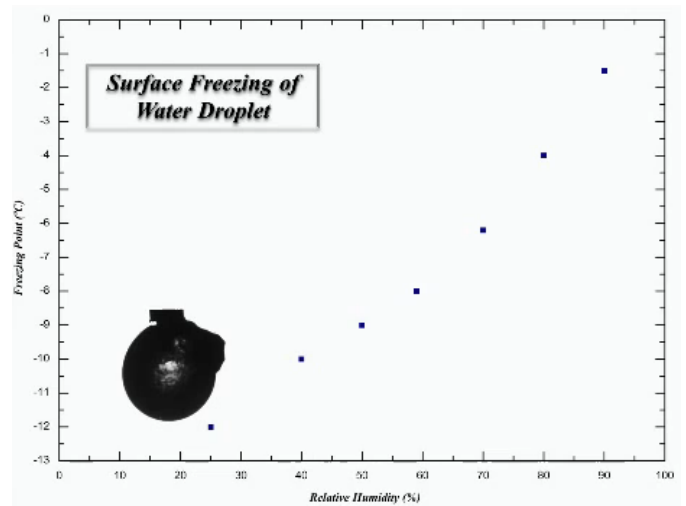

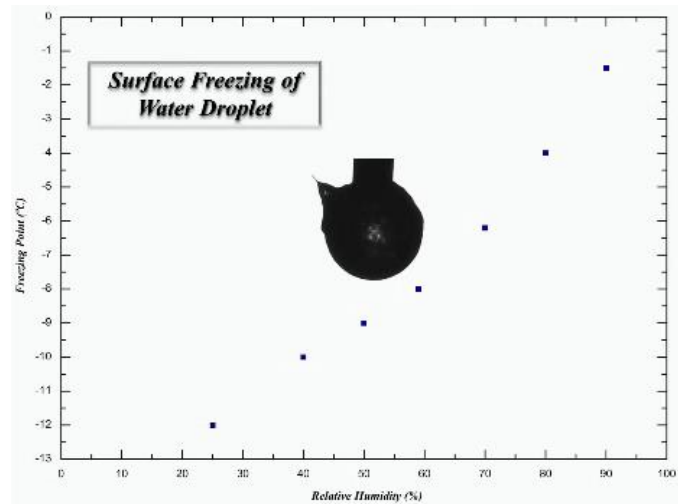

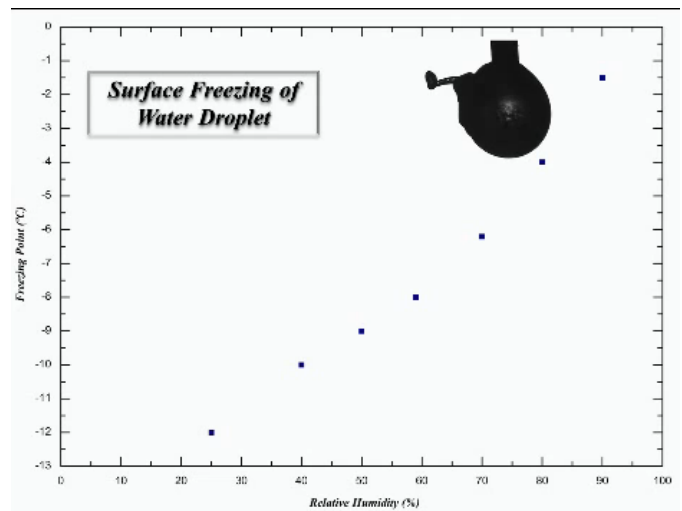

As soon as it flows out the outter shell freezes making a conduit for the pressed liquid water that flows through it. This method of growth requires the existence of a liquid tip and a simultaneous axial progression of the freezing wavefront all around the cylinder (faster than that of bulk).

The size and shape of these protruding icicles strongly depends on the humidity of air and subsequently on temperature, as can be seen in Figure 3 above. The lower the temperature is, the faster the freezing wavefront in bulk. For low humidity, the freezing temperature is also low and there are only protruding deformations. For medium humidity, the protuberances become conic and even present a final point. Finally, for high humidity, the freezing temperature approaches 0ºC and icicles protrude, forming stable cylindrical shells in which liquid water is contained and flows to the tip where they grow.

Experimental freezing points are repetitively measured with very small deviation at a unique temperature for a fixed humidity. No freezing occurs for temperature above that point. This indicates that freezing temperatures shown in Figure 2 correspond to points for which $\Delta G_S = 0$ where $\Delta G_S$ is the the variation of surface Gibbs free energy that occurs in case of freezing. At constant temperature T, according to our experimental results $\Delta G_S$ must depend on humidity. Moreover it will be positive (non-spontaneous) at low humidity and negative (spontaneous) for high humidity. In fact, we can write $\Delta G_S = \Delta H_S - T\Delta S_S$ and therefore it could only be explained by a decrease of $\Delta H_S$ with humidity or an increase of $\Delta S_S$ with humidity

$\Delta H_S$ is the change of surface enthalpy, which in our case can be identified with the change of surface energy or surface tension and $\Delta S_S$ is the change of surface entropy. For supercooled water at a certain $T$, $H_S$ is almost constant under variations of the relative humidity as can be seen in Figure 1. Assuming surface tension of both ice-air and liquid water-ice at that same $T$ to be also independent of the humidity of air, we can say that it is the jump of surface entropy which has to depend on humidity. The variation of ice surface entropy with respect to relative humidity (in %) can be estimated as $2.7 \times 10^{-8} \frac{N}{m \cdot K}$ approximately. This could be interpreted as follows.

Water molecules from the gas phase colliding against the surface of ice interact with those molecules on the surface, which are highly ordered in a crystalline solid phase. Non-polar nitrogen or oxygen molecules are not as effective disturbers of water molecules as water molecules themselves are. Therefore, for constant pressure, a higher humidity in air can be associated with a higher proportion of water molecules colliding against the surface of ice and disturbing the order of the molecules on it. This makes the surface entropy on the ice surface increase with the relative humidity of air.

Additionally, molecules on the surface of liquid water are not ordered in a lattice, but move in relative disorder. The effect of collisions of water molecules coming from the gas phase seems to be much smaller for the liquid than for the solid. In other words, gas water molecules are effective at disordering ordered solid molecules on ice but they cannot disorder disordered liquid molecules on supercooled water.

Finally, for constant temperature and pressure, the increase of surface entropy on ice with humidity makes $\Delta G_S$ for freezing decrease with humidity. For example, as can be seen in Figure 2, at -8ºC the droplet remains liquid whenever humidity is lower than 60%. It freezes when humidity is higher than that.

It is to be noted that so far it has been commonly assumed that dryness in air enhances evaporation on the surface, cooling the surface and making freezing more likely to occur at higher temperatures[10,17]. Here we experimentally demonstrate just the opposite: freezing nucleated on the surface of a water droplet is strongly depressed by dryness in air.

In order to test whether the evaporation on the surface could be strong enough to lower the temperature of the liquid we used a thermocouple wetted by a water droplet and a second thermocouple in air not far from the first one. A difference lower than 0.04ºC was registered in all cases. We can say that freshening by evaporation is overwhelmed by heat transport mechanisms like conduction and convection.

On the contrary, the disordering effect of water molecules in the gas phase colliding against and interacting with water molecules on the surface of the ice crystals is higher producing the strong depletion of the freezing point with dryness of the air.

This should shed light not only on water-related effects but also on surface catalysed reactions or changes of phase. It should be noted that it is the composition of the gas phase "A" that triggers a reaction in a bulk "B" by increasing the interface entropy of the final state of "B". It is not necessary a variation of surface properties of the initial state.

Moreover, the perplexing Mpemba effect[18] can easily be explained by our findings under certain circumstances as well. This effect is simply that very hot water (circa 100ºC) in a bowl freezes in a shorter time than cold water when they are both put in a freezer. After our results the Mpemba effect can be explained as follows. Hot water abundantly evaporates and humidifies the air in the freezer. Cold water evaporates insufficiently to saturate the air. Therefore, while the system's hot water-air cools down and keeps air saturated, the cold water-air cools down and keeps air dry. Therefore hot water-humid air only has to reach 0ºC to freeze, whereas cold water-dry air has to reach a lower temperature. The lower the temperature is, the exponentially slower the cooling. It therefore takes a much longer time to cool down cold water under 0ºC than to cool down hot water at the beginning. In this context the hot water takes a shortcut, as it were, and freezes faster than the cold water.

## Conclusions

Freezing of water droplets in still air is demonstrated to be triggered on the very surface. As an interface phenomenon it is influenced by properties of both media –particularly humidity in air-.

In spite that so far it had been assumed that dryness should rise up the Freezing point of water droplets, our experiments demonstrate that dryness strongly lowers it down. For instance a Freezing point of -12ºC has been repeatedly measured for 25% RH.

Moreover, surface freezing is conditioned not only by the properties of initial-state phases (liquid-gas) but mainly by the final-state phases (solid-gas). In this case, it is shown that Surface Entropy of the Ice-air interface increases with humidity in air and becomes determinant for freezing.

## Acknowledgements
This work has been supported by MAGSOAR SL. We thank E. Diez and I. Valiente for their cooperation and constructive comments. We also thank F. Serrano for technical support.

## Author contribution

J.-L.P.D. Conceived the original idea of this research work, analysed data and proposed the physical interpretation.
M.A.A.V. designed and built the hardware and collected and processed data.
J.S.G.C. cooperated in the first experimental tests and the design of electronics.
S.J.L. carried out the sets of experiments.

## Note

The method for freezing or supercooling without freezing here described was previously claimed in WIPO Patent Application WO/2015/107190.

## Links to video files

Ice Surface Entropy Induction by Humidity or How Humidity Prompts Freezing - icicle HR 25%
https://youtu.be/jd8m7dbAR14

Ice Surface Entropy Induction by Humidity or How Humidity Prompts Freezing - icicle HR 60%
https://youtu.be/DgCeWztCWCY

Ice Surface Entropy Induction by Humidity or How Humidity Prompts Freezing - icicle HR 80%
https://youtu.be/uj_UXhyhL6Y